# On Quantitatively Measuring Controllability of Complex Networks

CAI, Ning[1, 2]

[1]College of Electrical Engineering, Northwest University for Nationalities, Lanzhou, China
[2]School of Automation, Hangzhou Dianzi University, Hangzhou, China
E-mail: caining91@tsinghua.org.cn

**Abstract:** This letter deals with the controllability issue of complex networks. An index is chosen to quantitatively measure the extent of controllability of given network. The effect of this index is observed based on empirical studies on various classes of network topologies, such as random network, small-world network, and scale-free network.

## 1. Introduction

The controllability of a dynamic system reflects the ability of influencing from external input information to the motion of the overall system. Observability and controllability are dual alternatives. Integrated with stability, they form the theoretical foundation for most of the systems analysis and synthesis problems. Therefore, controllability has been one of the most important concepts in modern control theory.

Since about a decade ago, the controllability problems of dynamic networked large-scale systems have intrigued many scholars from both the control [1-7] and the physics [8-13] community, and will surely continue to attract the attention from more and more disciplines.

Tanner [1] earlier studied the controllability of systems with single leader and conjectured that excessive connectivity might even be detrimental to controllability, while giving a definition of graph controllability based on a partition of the associated Laplacian matrix. Paying attention to the relationship between the extent of symmetry of graph and controllability, Rahmani and Mesbahi [2] further extended the results in [1]. Cai *et al.* addressed the controllability problems of a class of high-order systems, proposing a scheme of controllability improvement [3-4]. Liu *et al.* [5] concerned the controllability of discrete-time systems with switching graph topologies. Ji *et al.* [6-7] dealt with the interactive protocols, endeavoring to integrate the influence of three facets upon controllability, which are the protocol, the vertex dynamics and the network topology, respectively.

Liu *et al.* [8] addressed the structural controllability of complex networks. They selected an index denoted by $N_D$ to quantitatively measure the extent of controllability of complex networks, namely the least amount of independent input signals required. Along the route of [8], there have emerged plenty of researches from the physics community, e.g. [9-13]. Particularly, Yan *et al.* [9] concentrated upon the problem of minimal energy cost for maneuvering the nodes. Yuan *et al.* concerned the exact controllability [10] of undirected networks with identical edge weights and discovered certain consistency with [8].

The concept of controllability for dynamic systems was originally raised by Kalman, with a set of algebraic criteria to check whether or not a given system is exactly controllable. It has formed the foundation of the controllability theory. However, there are two intrinsic problems limiting the study of the controllability of complex networks from the viewpoint of exact controllability. The first problem is that almost any arbitrary system is controllable in the sense of exact controllability. This fact reduces the significance of being controllable. As the second problem, it is rather difficult to translate those algebraic criteria into straightforward conditions for the topologies of networks.

In comparison with exact controllability, the concept of structural controllability possesses some advantages for coping with the controllability problems of networked systems. First, it bears no ambiguity like almost exact controllability. Being structurally controllable or not is essentially distinct for any system. Second, it is possible to acquire concise and straightforward criteria about the topologies of networks to check whether or not they are structurally controllable. Nevertheless, the essence of structural controllability is a minimal requirement for the availability of input information across the network, which is only a necessary prerequisite for controlling, whereas it cannot facilitate to evaluate the efficiency of control. Therefore, structural controllability is also restrictive.

In the current letter, a third angle on controllability is addressed, other than the exact controllability and structural controllability. We shall endeavor to explore the possible way to measure the extent of controllability of any given network, quantitatively. It is motivated by a wish to overcome some of the limitations of exact controllability and structural controllability, and to extend the methodology for controllability analysis of networks, from qualitative to quantitative. An index will be proposed for evaluating the ability of dynamic network to be easily controllable via the input information, and simulations will be performed on three distinct types of complex networks, namely the E-R networks, the WS small-world networks, and the scale-free networks, to illustrate the diversity of controllability situations.

The remaining part of this letter is organized as follows. Section 2 introduces the fundamental preliminaries about controllability of complex networks and describes the model concerned. Section 3 endeavors to analyze the computational controllability concept theoretically, based on some simple examples. Section 4 presents a series of experimental results toward three typical complex networks. Finally, Section 5 concludes this letter.

## 2. Problem Description and Preliminaries

### 2.1 Exact Controllability

The dynamics of a network concerned with $N_f$ followers and $N_l$ leaders is described by the following equation:

$$\dot{\xi} = A_{ff}\xi + A_{fl}\eta \tag{1}$$

where the vector $\xi(t) \in R^{N_f}$ represents the states of the follower vertices; $\eta(t) \in R^{N_l}$ represents the states of the leader vertices, which can be determined externally; and $A \in R^{N \times N}$ is the weighted adjacency matrix of the overall network topology, which is decomposed as follows according to the leader-follower distribution

$$A = \begin{bmatrix} A_{ff} & A_{fl} \\ A_{lf} & A_{ll} \end{bmatrix}$$

The network is exactly controllable if the state values of the followers can be fully controlled via appropriately selected state values of leaders, otherwise, it is uncontrollable. This conforms to the conventional definition of Kalman controllability.

*Definition 1:* A dynamic network (1) is exactly controllable if for any initial values of follower states

$$\xi_1(0), \xi_2(0), ..., \xi_{N_f}(0) \in R$$

there exist $\tau < \infty$ and proper leader signals

$$\eta_1(t), \eta_2(t), ..., \eta_{N_l}(t) \quad (t \in [0, \tau])$$

such that $\xi_1(\tau) = \xi_2(\tau) = ... = \xi_{N_l}(\tau) = 0$.

Lemma 1 below provides the most fundamental criterion to check controllability, known as the rank test.

*Lemma 1 [14]:* A dynamic network (1) is exactly controllable if and only if the controllability matrix is of full rank, which is

$$\Psi = \begin{bmatrix} A_{fl} & A_{ff}A_{fl} & A_{ff}^2 A_{fl} & \cdots & A_{ff}^{N_f -1} A_{fl} \end{bmatrix}$$

*Remark 1:* If the number of leader $N_l = 1$, then the controllability matrix $\Psi$ is square. In this case, the network is controllable if and only if $\Psi$ is nonsingular.

## 2.2 Almost Exact Controllability

The classic notion of exact controllability of dynamic systems is qualitative, and could answer only "Yes" or "No". However, it is well-known by the control theory community that almost any linear system is controllable, i.e. the probability for any system (1) to be exactly controllable is 1, so long as the matrix *A* is continuous-valued; whilst the uncontrollable cases are just the rare exceptions [4, 15-16]. In this sense, the exact controllability is less significant, whereas a better way to evaluate the extent of controllability should be to measure how far a given network is from being uncontrollable [15-16]. Such a route may be called computational controllability.

The main connotation behind the concept of computational controllability can be illustrated by Fig. 1.

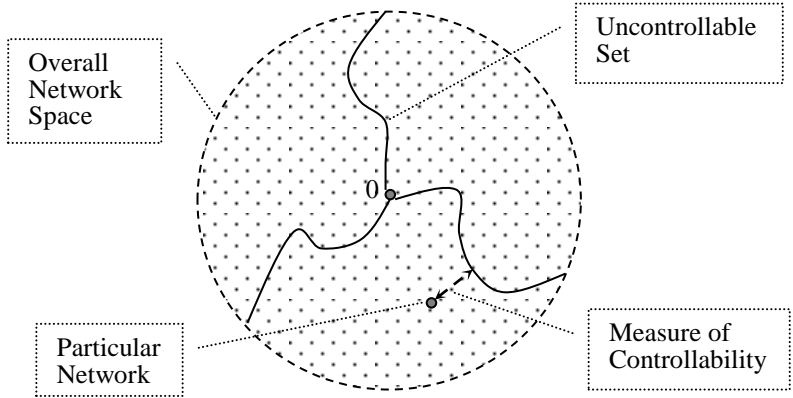

Fig. 1. Network space of $R^{N \times N}$.
Solid curves represent the set of networks that cannot be exactly controlled.

In Fig. 1, any network is represented by a point in the space. Almost all the networks are exactly controllable. If a particular network happens to locate at the uncontrollable set, then any infinitesimal perturbation could let it leave this set and become controllable.

## 3. Computational Controllability of Complex Networks: Theoretical Analysis

### 3.1 Introduction of Concept

Almost all the dynamic networks are controllable. Nevertheless, the relative controllability should be different for various network topologies. Some may possess relatively higher controllability, whereas some others may be less controllable as being very close to an uncontrollable network.

According to Lemma 1, controllability is determined by the rank of $\Psi$. If the rank is nearly reduced, then the controllability of the associated network should be low. If $\Psi$ is square as the single leader case, it is expected to be away from being singular for higher controllability. Thus, it should be meaningful to measure the extent of its nonsingularity. In this paper, we choose the conditional number of $\Psi$ as this measure.

*Definition 2:* The conditional number of a square matrix $\Psi$ is

$$cond(\Psi) = \|\Psi\| \|\Psi^{-1}\|$$

Conditional number can be regarded as a measure of nonsingularity, which varies within $[1,\infty]$. $cond(\Psi)=1$ if $\Psi$ is identity matrix, with the maximal possible nonsingularity. Oppositely, if $\Psi$ is nearly singular, then $cond(\Psi)$ must be large because $\|\Psi^{-1}\|$ is large. The greater the conditional number, the closer the matrix $\Psi$ is to be singular, till $cond(\Psi)=\infty$ as the rank of $\Psi$ becomes reduced. A definition of computational controllability hereby arises, based on conditional number of controllability matrix.

*Definition 3:* The computational controllability of a dynamic complex network (1) can be measured by

$$\kappa = \frac{1}{cond(\Psi)}$$

where $\Psi$ is the controllability matrix.

*Remark 2:* For any network, $\kappa \in [0,1]$. The greater this value, the higher controllability a network achieves. A network is uncontrollable if $\kappa = 0$, otherwise, it is controllable. However, if $\kappa$ is very small, the controllability is still quite weak although the network is thought to be controllable in the conventional sense.

### 3.2 Basic Analysis

Let us start the theoretical analysis on computational controllability from investigating several concrete typical instances.

*Example 1:* Consider the path $P_N$. For simplicity, let $N=5$ and let the fifth agent be the only leader. It is always controllable in the sense of exact controllability. The network topology is illustrated in Fig. 2.

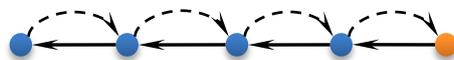

Fig. 2. Fifth order path $P_5$ with default arc weight 1.

If this network is directed, its adjacency matrix is

$$A = \left[\begin{array}{cccc:c} 0 & 1 & 0 & 0 & 0 \\ 0 & 0 & 1 & 0 & 0 \\ 0 & 0 & 0 & 1 & 0 \\ 0 & 0 & 0 & 0 & 1 \\ \hdashline 0 & 0 & 0 & 0 & 0 \end{array}\right]$$

which is already decomposed by dotted lines. The controllability matrix $\Psi$ is just the identity matrix, with $\kappa = 1$, indicating a maximal controllability.

If this network is undirected with the adjacency matrix

$$A = \begin{bmatrix} 0 & 1 & 0 & 0 & 0 \\ 1 & 0 & 1 & 0 & 0 \\ 0 & 1 & 0 & 1 & 0 \\ 0 & 0 & 1 & 0 & 1 \\ \hdashline 0 & 0 & 0 & 1 & 0 \end{bmatrix}$$

the controllability matrix is

$$\Psi = \begin{bmatrix} 0 & 0 & 0 & 1 \\ 0 & 0 & 1 & 0 \\ 0 & 1 & 0 & 2 \\ 1 & 0 & 1 & 0 \end{bmatrix}$$

The nonsingularity of this matrix is also quite satisfactory with $cond(\Psi) = 5.8284$ and $\kappa = 0.1716$, indicating a rather high controllability.

*Example 2:* Consider the complete network $C_5$. The adjacency matrix is

$$A = \begin{bmatrix} 0 & 1 & 1 & 1 & 1 \\ 1 & 0 & 1 & 1 & 1 \\ 1 & 1 & 0 & 1 & 1 \\ 1 & 1 & 1 & 0 & 1 \\ \hdashline 1 & 1 & 1 & 1 & 0 \end{bmatrix}$$

with the corresponding controllability matrix being

$$\Psi = \begin{bmatrix} 1 & 3 & 9 & 27 \\ 1 & 3 & 9 & 27 \\ 1 & 3 & 9 & 27 \\ 1 & 3 & 9 & 27 \end{bmatrix}$$

It is singular with the rank being only 1. Thus, $\kappa$ is definitely zero.

*Example 3:* Consider the random weighted directed network, which is known to be almost controllable in the conventional sense. A random network of 8th order is generated, with the last agent assigned as the leader. The resulted controllability matrix is

$$\Psi = \begin{bmatrix} 0.1897 & 0.8962 & 2.7716 & 8.6052 & 27.0802 & 84.4664 & 264.5718 \\ 0.1934 & 1.9397 & 5.3306 & 17.1429 & 53.1950 & 166.8824 & 521.6511 \\ 0.6822 & 1.0615 & 3.9149 & 12.0033 & 37.6166 & 117.6977 & 368.1719 \\ 0.3028 & 1.0436 & 2.9809 & 9.6179 & 29.8652 & 93.5949 & 292.6930 \\ 0.5417 & 1.1522 & 3.9331 & 11.9725 & 37.7541 & 117.8886 & 368.9820 \\ 0.1509 & 1.5067 & 3.9574 & 12.9025 & 39.9819 & 125.3981 & 392.0307 \\ 0.6979 & 1.0432 & 4.1353 & 12.2815 & 38.9403 & 121.4206 & 380.1693 \end{bmatrix} \quad (2)$$

In this case, $cond(\Psi) = 1.5075 \times 10^6$, and $\kappa = 6.6335 \times 10^{-7}$. Such a network obviously holds a lower controllability.

The controllability of a network is closely relational to its spectral distribution, by which the reason for lower controllability of Example 3 could partially be explained.

Note that the average values of column elements keep on increasing from left to right in (2). Such a phenomenon will greatly deteriorate the nonsingularity of the overall matrix, according to the following lemma.

*Lemma 2:* The infinite conditional number of square matrix $A = [a_{ij}] \in R^{N \times N}$ is generally greater than $\gamma N$, with $\gamma \geq 1$ being the ratio between the maximal and minimal row/column sum of absolute values of elements.

*Proof:* If $A$ is singular, $cond(A)_\infty$ is infinite. Consider the case that $A$ is nonsingular. Without loss of generality, let the rows be sorted according to the sums of absolute values of elements, with the first row possessing the minimal sum, and

$$|a_{11}| = \frac{1}{N} \sum_{j=1}^{N} |a_{1j}|$$

Now consider the matrix $A/|a_{11}|$, which has obviously the same conditional number with $A$. Because the sum of absolute values of elements in the first row of $A/|a_{11}|$ is

$$\sum_{j=1}^{N} (|a_{1j}|/|a_{11}|) = N$$

the sum of absolute values of elements in the last row of $A/|a_{11}|$ is $\gamma N$ and as a result

$$\|A/|a_{11}|\|_\infty = N\gamma$$

The inverse of matrix $A/|a_{11}|$ can be computed through a series of elementary row transformations to the following augmented matrix:

$$\begin{bmatrix} a_{11}/|a_{11}| & a_{12}/|a_{11}| & \cdots & a_{1N}/|a_{11}| & 1 & 0 & \cdots & 0 \\ a_{21}/|a_{11}| & a_{22}/|a_{11}| & \cdots & a_{2N}/|a_{11}| & 0 & 1 & \cdots & 0 \\ \vdots & \vdots & & \vdots & & & \ddots & \\ a_{N1}/|a_{11}| & a_{N2}/|a_{11}| & \cdots & a_{NN}/|a_{11}| & 0 & 0 & \cdots & 1 \end{bmatrix}$$

Evidently, the absolute value of the element of $(A/|a_{11}|)^{-1}$ with index (1, 1) is 1, and it follows that

$$\left\| (A/|a_{11}|)^{-1} \right\|_{\infty} > 1 \quad \square$$

*Remark 3:* The bound $\gamma N$ given in Lemma 2 is quite conservative. Usually, the conditional number of a matrix is far greater than this bound. Besides, although there are different definitions of conditional number, generally their trends are consistent. It is not possible that the conditional number is very large by one definition but very small by another.

The spectrum of $A_{ff}$ in Example 3 is

$$\{3.1285, -0.4372 \pm 0.445i, -0.3542 \pm 0.2556i, -0.7864, -0.7592\}$$

Such a distribution is typical for the class of random networks in Example 3, with one largest eigenvalue and the remaining being very smaller.

It is known that any random matrix is almost diagnolizable [17]. Suppose that $A_{ff}$ could be diagnolized as

$$P^{-1} A_{ff} P = \begin{bmatrix} \lambda_1 & & \\ & \ddots & \\ & & \lambda_{N_f} \end{bmatrix}$$

It follows that

$$A_{ff}^k = P \begin{bmatrix} \lambda_1^k & & \\ & \ddots & \\ & & \lambda_{N_f}^k \end{bmatrix} P^{-1}$$

The magnitude of $A_{ff}^k$ of Example 3 increases with $k$ because $\lambda_1$ is much larger than 1.

The controllability can be changed by scaling the spectral radius through zooming the topology of network, without affecting the structure.

Consider the controllability of $A/\mu$, where $\mu > 0$ is the zoom factor. Take the data in Example 3 and let the zoom factor $\mu$ vary across $[1, 4]$; the corresponding controllability variation forms a quite smooth curve, as shown in Fig. 3. One can see that, at first, the controllability increases as $\mu$ increases, and it reaches a maximum when $\mu = 2.4$; then the controllability keeps on decreasing because the magnitude of $A_{ff}^k$ decreases with $k$ if $\lambda_1$ is less than 1.

Intuitively speaking, the controllability of a network should generally be better if the eigenvalues locate nearby the unit circle. The spectral radius being either greatly larger or less than 1 will deteriorate the controllability.

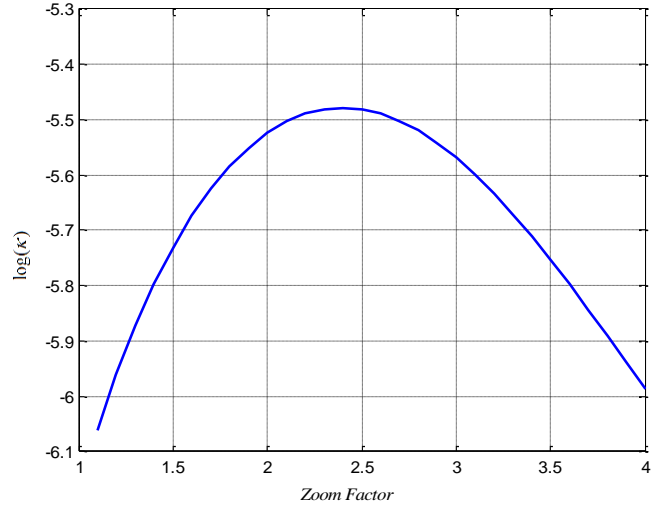

Fig. 3. Trend of controllability in Example 3 with zoom factor varying across [1, 4]

## 4. Computational Controllability of Complex Networks: Empirical Analysis

This section will manifest a series of empirical analysis on computational controllability for the three typical classes of complex networks.

### 4.1 E-R Network

Consider the E-R networks comprising of 15 followers and 1 leader. Let the connectivity probability $p$ vary across [0, 0.9]. The value of each data point is the average of 2000 random trials. The variation of controllability is shown in Fig. 4. One can see that the logarithm variation trend could be fitted into two straight lines.

It is worth mentioning that throughout this section, the edge weights are mounted with noises uniformly distributed within [-0.025, 0.025], which are relatively rather weak.

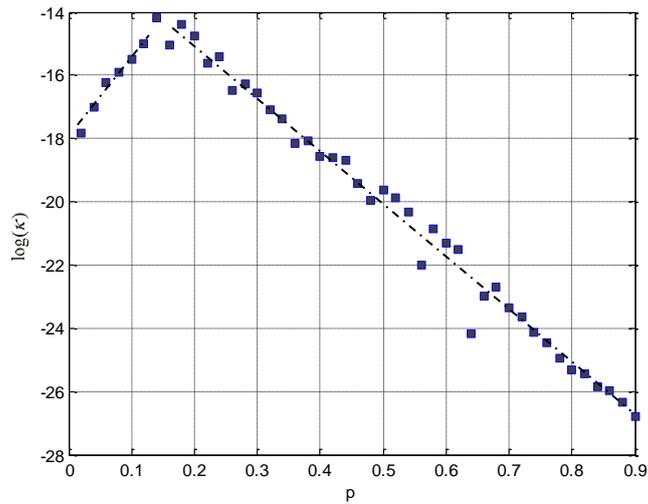

Fig. 4. Trend of controllability of E-R network with connectivity probability varying across [0, 0.9]

The controllability achieves a maximum when about $p = 0.15$. If the value of $p$ is too low, the network lacks sufficient connectivity to guarantee controllability. However, the controllability keeps on decreasing as $p$ is large enough. The more a network approximates a complete graph, the less controllable it would be.

Now fix the connectivity probability $p$ to 0.15 and let the number of leaders $N_l$ vary across 1, 2, …, 15. The variation of controllability is shown in Fig. 5. There is a positive correlation between the controllability and $N_l$. However, such a correlation becomes weak as the number of leaders is sufficiently large. For this example, $\log(\kappa)$ seems to approach a limit being less than 10.

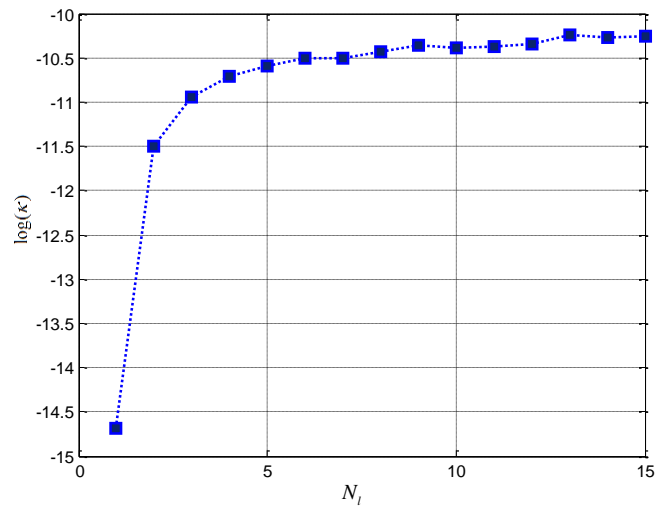

Fig. 6. Trend of controllability of E-R network with number of leaders varying across 1, 2, …, 15

Fix the connectivity probability $p$ to 0.15 and the number of leaders $N_l$ to 1. Let the number of followers vary across 5, 6, …, 104, 105. According to the result, the controllability is negatively correlated to the number of followers, as shown in Fig. 6. One can see that the controllability of a network with large-scale is quite weak.

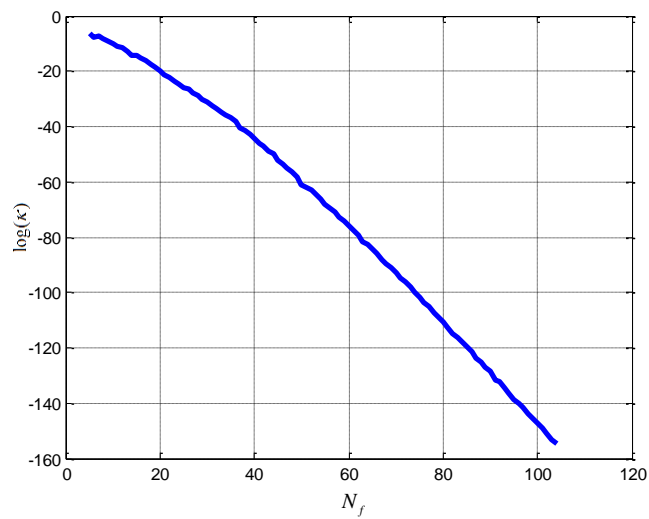

Fig. 7. Trend of controllability of E-R network with number of followers varying across 5, 6, …, 104, 105

**4.2 WS Small-World Network**

Consider the WS small-world networks containing 16 followers, with $K=2$. First, let the switching probability $p$ vary across [0, 0.9]. However, no direct relationship can be seen between the controllability and $p$, according to the simulation result.

Set the switching probability $p$ to 0.5 and successively increase the quantity of leaders $N_l$ from 1 to 15, with the quantity of followers $N_f$ being fixed. A resulted relationship between the controllability and $N_l$ is shown in Fig. 8. It can be seen that more leaders are generally beneficial to the controllability, although this effect is less prominent if $N_l \gg 1$.

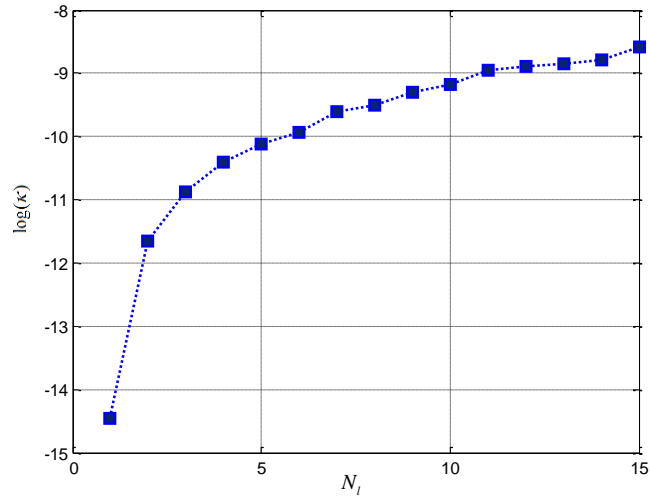

Fig. 8. Trend of controllability of WS small-world network with number of leaders varying across 1, 2, …, 15

Let $p=0.5$, $N_l=1$, and the quantity of followers be 5, 6, …, 85. The relationship between the controllability and $N_f$ is illustrated in Fig. 9, which may be approximately fitted into a straight line.

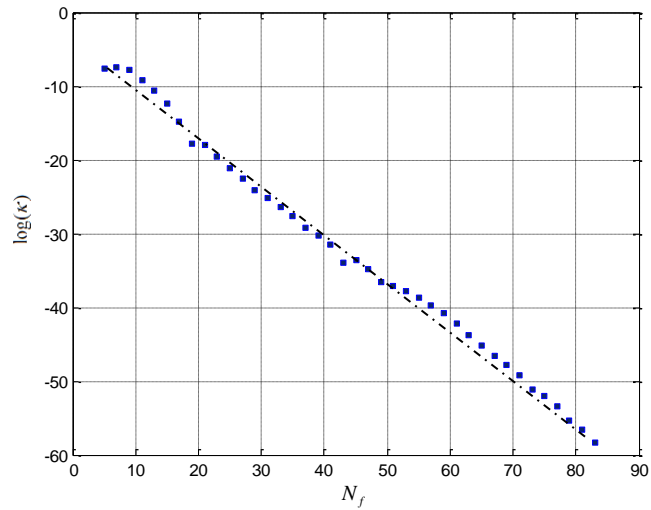

Fig. 9. Trend of controllability of WS small-world network with number of followers varying across 5, 6, …, 85

Let $p=0.5$, $N_l=1$, and $N_f=30$; successively increase the $K$ from 2 to 15. It should be interesting to see that although being independent of *p*, the controllability is negatively correlated to *K*, as shown in Fig. 10.

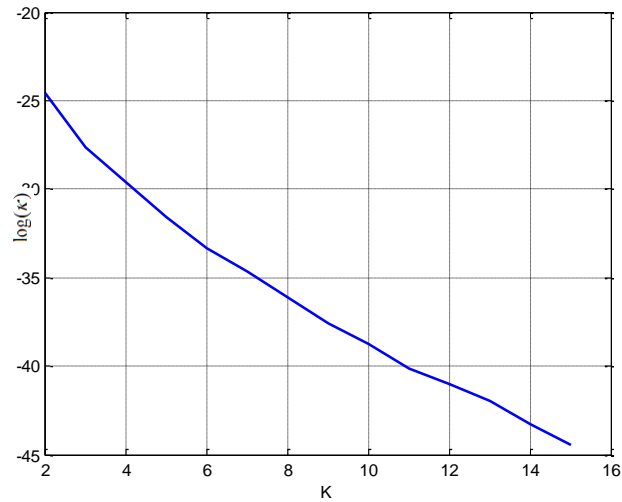

Fig. 10. Trend of controllability of WS small-world network with *K* varying
across 2, 3, …, 15

### 4.3 BA Scale-Free Network

Consider the controllability of BA scale-free networks. First, let the order of the original complete graph $m_0 = 7$, the number of added vertices $t = 8$, and the number of new edges per vertex $m = 3$. The resulted network topology is illustrated in Fig. 11, with the indices of vertices sorted by degree.

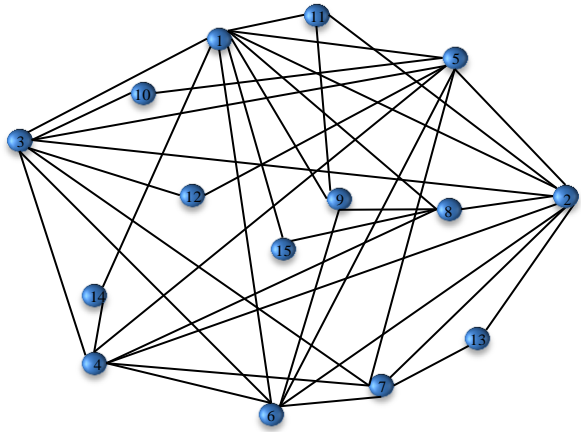

Fig. 11. BA scale-free network with $m_0 = 7$, $m = 3$, and $t = 8$

Select each vertex as the single leader, one by one, with the computed controllability of network successively shown in Fig. 12. A slight tendency of positive correlation between the controllability and the degree of leader vertex can be observed.

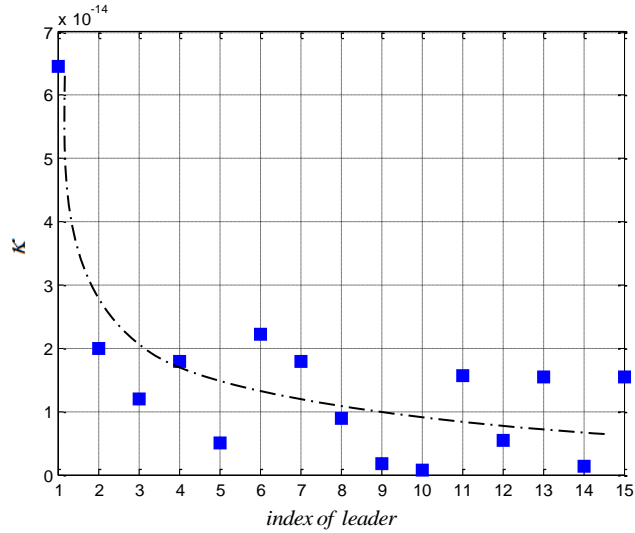

Fig. 12. Trend of controllability of BA scale-free network with degree of leader decreasing

Fix $m_0 = 7$ and $t = 8$; meanwhile, let $m$ vary across 1, 2, …, 6. For each case, the average controllability of 2000 trials is recorded and displayed in Fig. 13. Evidently, the controllability is positively correlated to $m$.

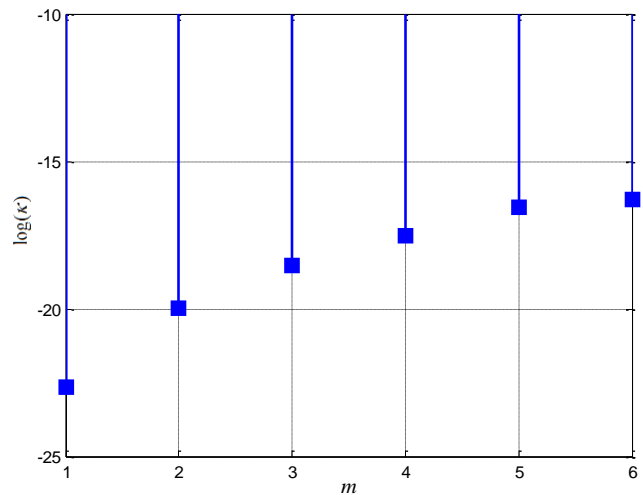

Fig. 13. Trend of controllability of BA scale-free network with $m$ varying across 1, 2, …, 6

## 5. Conclusion

Considering the fact that weighted real-world networks are almost controllable in the sense of exact controllability, this letter addresses the problem of measuring the controllability of complex networks quantitatively. For this end, an index is selected, based on the conditional number of controllability matrix. The effect of this index is observed and discussed, mainly by a series of experiments on various types of complex networks, e.g. the E-R networks, the WS small-world networks, and the BA scale-free networks.

Studies along this route could enable the comparisons of controllability between complex networks with different topologies. The current letter just takes a first step. There exist abundant directions of potential future extensions, for instance: 1) Other indices beside the one here can be sought; 2) More experiments can be conducted to a broader range of networks, for the sake of discovering more empirical regularities; 3) Certain phenomena observed could be explained analytically. 4) The network model concerned can take other forms, e.g. being represented by Laplacian matrix.

## Acknowledgments

This work is supported by National Natural Science Foundation (NNSF) of China (Grants 61263002 & 61374054), by Program for Young Talents of State Ethnic Affairs Commission (SEAC) of China (Grant [2013] 231), and by the Zhejiang Open Foundation of the Most Important Subjects.